\documentclass[reprint,aps,prl,twocolumn,superscriptaddress,amsmath,amssymb,floatfix,footinbib,longbibliography,notitlepage]{revtex4-1}
\usepackage{graphicx}
\usepackage{xcolor}
\usepackage{bm}
\usepackage[colorlinks, linkcolor= blue, citecolor = blue, urlcolor=blue]{hyperref}
\usepackage{ulem} 
\usepackage{physics}
\usepackage{float}
\usepackage{wrapfig}
\usepackage[stable]{footmisc}
\usepackage{array}
\usepackage{multirow}
\usepackage{hhline}
\usepackage{pifont}
\usepackage{chemformula}
\usepackage{textcase} 
\usepackage{soul}

\def\nn{\nonumber}
\def\bea{\begin{eqnarray}}
\def\eea{\end{eqnarray}}
\def\be{\begin{equation}}
\def\ee{\end{equation}}

\def\bal{\begin{aligned}}
\def\eal{\end{aligned}}

\usepackage{babel}

\begin{document}
\title{Disorder-driven Weyl-Kondo Semimetal Phase in WTe$_2$}

\author{Arpan Manna}
\author{Sunit Das}
\author{Amit Agarwal}
\email{amitag@iitk.ac.in}
\author{Soumik Mukhopadhyay}%
\email{soumikm@iitk.ac.in}
\affiliation{Department of Physics, Indian Institute of Technology Kanpur, Kanpur 208016, India}

\begin{abstract} 

In this Letter, we report the observation of disorder-driven anisotropic Kondo screening and spontaneous Hall effect in bulk WTe$_2$, a nonmagnetic type-II Weyl semimetal. We show that Kondo scattering emerges more prominently in disordered samples and produces magnetoresistance that is strongly anisotropic with respect to both current and magnetic field orientation, reflecting the underlying type-II Weyl dispersion. Strikingly, we find a spontaneous Hall effect in zero magnetic field, whose magnitude is enhanced with disorder, together with a large second-harmonic Hall signal exhibiting quadratic current scaling. Our analysis indicates that disorder-driven Kondo interactions pin the Fermi level near the Weyl nodes. This enhances the Berry curvature–driven nonequilibrium transport, accounting for both the second-order and spontaneous Hall responses. These findings establish disordered WTe$_2$ as a platform hosting Weyl–Kondo fermions and highlight disorder as an effective control knob for inducing correlated topological phases in weakly correlated Weyl semimetals.

\end{abstract}

\maketitle

{\it {Introduction:--}} The interplay between electronic correlations and band topology provides a fertile ground for emergent strongly correlated phases in quantum systems~\cite{Keimer_np17,Paschen_nrp20,Paschen_nrm24, Dzero2016, Chen_NP22,  Guo2023, Juyal_prb2017, Juyal_prl18, Juyal_prb22, Amit_prb18, Alapan_prb2025}. A paradigmatic example is the Weyl–Kondo semimetal (WKSM), in which Kondo screening renormalizes Weyl fermions into heavy quasiparticles~\cite{Laia}. This phase has been identified in non-centrosymmetric heavy-fermion compounds, notably the Ce$_3$Bi$_4$(Pt$_{1-x}$Pd$_x$)$_3$ family, where Weyl nodes are pinned near the Fermi level and acquire strongly renormalized velocities~\cite{Dzsaber, Dzsaber1, Dzsaber2, L.Chen, Dzsaber_NC22, Qimiao_pnas18}. WKSM behavior in such systems can be accessed through chemical substitution~\cite{Dzsaber, Dzsaber1}, pressure~\cite{Dzsaber_NC22}, magnetic fields~\cite{Lee_scd22}, or carrier doping~\cite{Fang_prr25}.  

In this Letter, we ask a central question: can disorder alone drive a Weyl–Kondo semimetal phase in a weakly correlated, nonmagnetic Weyl semimetal? We demonstrate that bulk WTe$_2$, a non-centrosymmetric, nonmagnetic, type-II Weyl semimetal of the transition-metal dichalcogenide family~\cite{Soluyanov_nature15, Y.Wang, ali2014large, L.Wang, Sante_prl17, Sharma_prb17, Lee_sc15}, can host a disorder-tuned Weyl–Kondo semimetal phase. In its orthorhombic $T_d$ phase, WTe$_2$ is a well-established type-II Weyl semimetal with low carrier density, and extremely large non-saturating magnetoresistance~\cite{ali2014large,cai2015drastic,thoutam2015temperature,zhu2015quantum,rhodes2015role,wang2016breakdown,fatemi2017magnetoresistance}, making it a good platform to probe disorder-driven correlated phases. 


We show that disorder tunes Kondo screening in WTe$_2$, producing anisotropic transport with respect to both current and magnetic field orientation, reflecting the underlying anisotropy of the type-II Weyl fermions~\cite{Soluyanov_nature15, L.J.Wang, J.Guimaraes}. 
More strikingly, we observe a disorder-enhanced spontaneous Hall effect in the absence of an external magnetic field~\cite{Dzsaber1, sur_25}, together with a sizable second-order Hall response driven by Berry curvature dipole~\cite{Fu_prl15, Pablo_NP18, Tewari_prb21, Sinha_NP22, Chakraborty_22, Kang_nm19, Kamal_prb23, Adak2024}. These results point to disorder-induced correlation effects that reproduce key signatures of a WKSM phase, where Kondo interactions dynamically pin the Fermi level near the Weyl nodes. Our findings establish disorder as a viable tuning parameter for realizing WKSM phases and opening new directions for exploring their transport and correlation-driven properties in other Weyl semimetals.

\begin{figure*}
\includegraphics[width=1\linewidth]{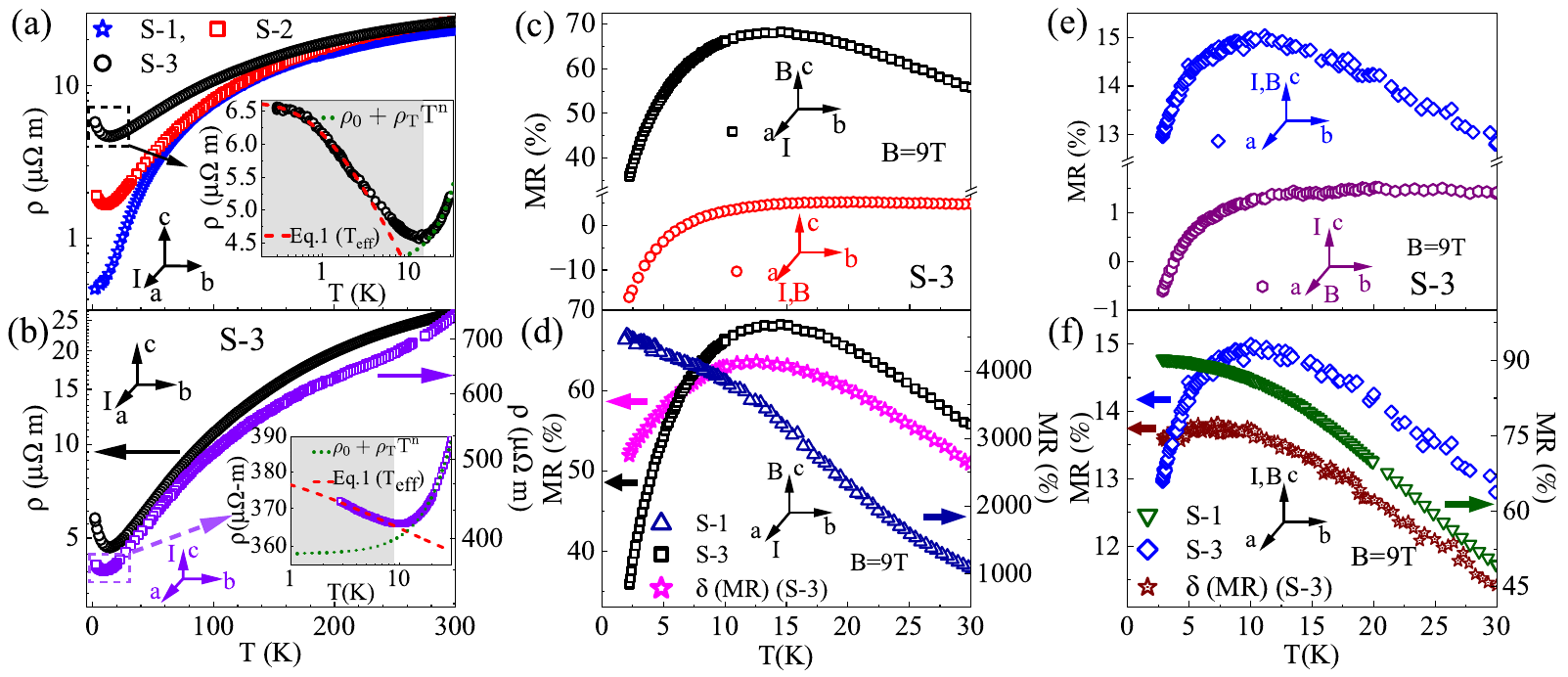}
\caption{{\bf Observation of anisotropic Kondo screening effect in $\mathrm{WTe_2}$.}
(a) Temperature dependence of the zero magnetic field longitudinal resistivity of S-1, S-2 and S-3 with current $\mathrm{(I)\parallel a}$. Inset: Low temperature resistivity of S-3 showing two different transport regimes with appropriate fits. The shaded area marked by $\mathrm{T<T_{K}}$ is the Kondo regime, with the resistivity fitted by  Eq.~\ref{eq1} ($\mathrm{T=T_{eff}}$), while the metallic regime is fitted by a power law equation (see main text).
(b) Temperature dependence of the zero magnetic field longitudinal resistivity of S-3 with $\mathrm{I\parallel a}$ and $\mathrm{I \parallel c}$, clearly showing anisotropic Kondo screening. Inset: The fits to the resistivity in the Kondo and the metallic regime, similar to the inset of (a).
(c) MR vs T at B = 9 T of S-3 ($\mathrm{I\parallel a}$) for $\mathrm{B\parallel c}$ and $\mathrm{B\parallel a}$.
(d) Temperature dependence of MR for S-3 (open squares) and S-1 (open triangles) with $\mathrm{I \parallel a}$ at B = 9 T. The open stars represent the difference [$\mathrm{\delta}$(MR)= MR ($\mathrm{I \parallel a}$, B $\parallel$ c) - MR ($\mathrm{I \parallel a}$, $\mathrm{B \parallel a}$)] for applied B = 9 T, depicting Kondo-contributed MR anisotropy with magnetic field direction. (e, f) Corresponding analysis for $\mathrm{I \parallel c}$, mirroring the procedures shown in (c) and (d). This quantifies the Kondo-contributed MR anisotropy for the out-of-plane current direction.}
\label{Fig1} 
\end{figure*}

{\it {Experimental methods:--}}  
Bulk single crystals of WTe$_2$ were synthesized by chemical vapor transport (CVT)~\cite{Linnartz}. To probe disorder effects systematically, we studied three representative samples, S-1, S-2, and S-3, with residual resistivity ratios (RRR) of $\sim$51, 15, and 6, respectively, for current along the $\rm a$-axis ($\rm I \parallel a$). The RRR values provide a quantitative measure of disorder strength, enabling a controlled comparison of disorder effects across samples.  

Structural quality was verified by X-ray diffraction (XRD), which confirmed $\rm c$-axis–oriented single-crystal growth. Elemental composition and stoichiometry were established using energy-dispersive spectroscopy (EDS) (see Supplemental Material (SM) ~\footnote{The Supplemental Materials includes detailed discussions on (S1) sample characterization through XRD, EDS and probing disorder via Shubnikov–de Haas oscillations, (S2) magnetic susceptibility, (S3) low temperature heat capacity, Table-1 containing fitting parameters of Eq.~(1) ($\rm T=T_{eff}$), (S4) temperature dependent resistivity at different magnetic field for different current and magnetic field directions, (S5) spontaneous Hall effect of S-1 and S-2, (S6) non saturating magnetoresistance, Hall resistivity at different temperatures and the two-band model analysis, (S7) calculation of spontaneous Hall effects in type-II Weyl semimetals.}, Sec.~S1). Magnetic susceptibility and heat capacity were measured in a Quantum Design PPMS (see SM, Sec.~S2–S3).  

Transport measurements were carried out using a standard six-probe lock-in technique ($f = 17.7$ Hz). Longitudinal ($\rm \rho_{xx}$) and Hall ($\rm \rho_{xy}$) resistivities were recorded with current and magnetic field applied along both the $\rm a$- and $\rm c$-axes. Nonlinear transport was probed via second-harmonic Hall measurements ($\rm \rho_{xy}^{2\omega}$), performed with current along the $\rm a$-axis to probe Berry curvature dipole-driven responses.

{\it{Anisotropic Kondo screening:--}}~Figure~\ref{Fig1}(a) shows the temperature-dependent resistivity of the three samples with $\rm I \parallel a$. While S-2 and S-3 exhibit a clear low-temperature upturn in resistivity, S-1 shows no such feature. At 2~K, the upturn magnitude reaches $\sim$25\% in S-3 and 16\% in S-2, decreasing systematically with increasing RRR. The systematic suppression of the upturn with increasing RRR suggests that Kondo scattering, arising from local magnetic moment formation, is significant in samples with lower RRR.
We show a direct comparison between in-plane ($\rm{I\parallel a}$) and out-of-plane ($\mathrm{I\parallel c}$) resistivity for S-3 in Fig.~\ref{Fig1}(b). The out-of-plane resistivity is almost two orders of magnitude larger than the in-plane resistivity, as expected for van der Waals layered materials. Remarkably, the resistivity upturn of the S-3 sample in the out-of-plane configuration is drastically reduced to just 1.7\%, compared to 25\% in the in-plane case. This pronounced directional dependence of the resistivity upturn is a signature of anisotropic Kondo screening in $\rm WTe_2$. 

\begin{figure*}[t]
\includegraphics[width=.8\linewidth]{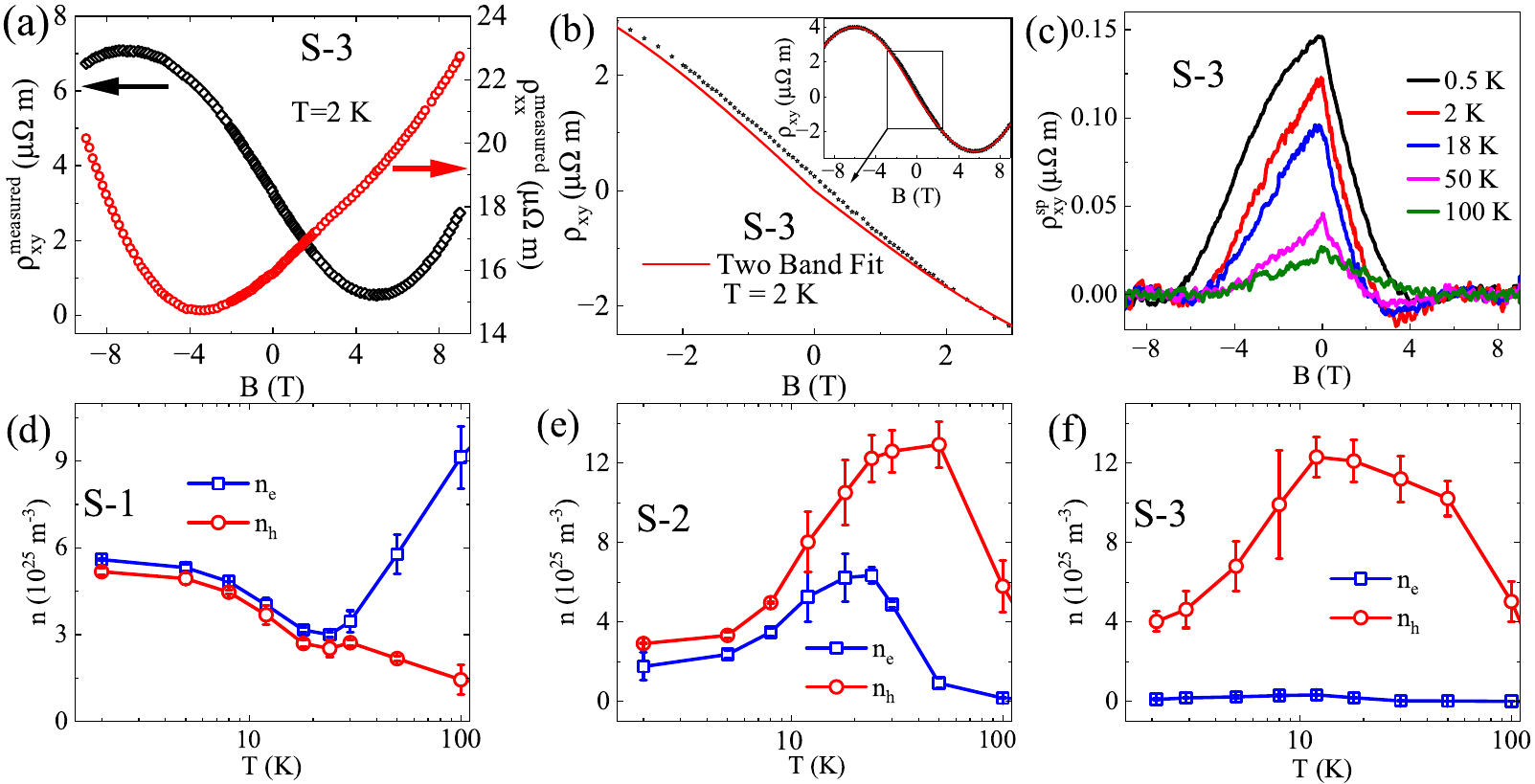}
\caption{{\bf Observation of spontaneous Hall effect and its co-relation with charge de-compensation.} (a) The measured Hall ($\mathrm{\rho^{measured}_{xy}(B)}$) and magneto resistivity ($\mathrm{\rho^{measured}_{xx}(B)}$) at 2 K showing influence of one on another due to contact misalignment. (b) The two-band model fit on the misalignment corrected Hall resistivity ($\mathrm{\rho_{xy}(B)}$) (full magnetic field range in the inset) shows a low-field anomaly presenting SHE. (c) Spontaneous Hall signal of S-3 at different temperatures obtained by subtracting the two-band model fitting from the Hall resistivity. The carrier concentrations of samples (d) S-1, (e) S-2, and (f) S-3 reflect the level of charge compensation and thus the shifting of the Fermi energy across these three distinct samples.}\label{Fig2}
\end{figure*}

The insets of Fig.~\ref{Fig1}(a) and (b) show the temperature dependence of the longitudinal resistivity of S-3 for $\rm I\parallel a$ and $\rm I\parallel c$, respectively.
Below the resistivity minimum (shaded region in the inset), the transport is dominated by Kondo scattering. Conduction electrons undergo spin-flip scattering from localized magnetic moments, producing a logarithmic increase in resistivity upon cooling. 
This crucial aspect of Kondo physics is well captured by the Hamann expression~\cite{D.R.Hamann, V.K.C.Liang, K.Fischer, J.Ge, Y.Wang.C.Xie, Bikash_prb2021},
\begin{equation}
\rm \rho(T) = \rho_H \left(1-\frac{\log(T/T_K)}{\sqrt{\log^2(T/T_K)+S(S+1)\pi^2}}\right),
\label{eq1}
\end{equation}
where $\rm \rho_H$ is a scaling constant, $\rm T_K$ the Kondo temperature, and $\rm S$ the effective spin quantum number. At higher temperatures, the resistivity is instead governed by a conventional power-law form ${\rm \rho(T)} = \rho_0 + \rm \rho_T \rm T^n$, with $\rm \rho_0$ the residual resistivity, $\rm \rho_T$ the temperature coefficient, and $\rm n$ is the power-law exponent.  

We observe that in a very low temperature regime where the Kondo divergence saturates, Eq.~\eqref{eq1} does not give a fully satisfactory fit to the resistivity. The observed minor discrepancies are attributed to the Ruderman–Kittel–Kasuya–Yosida (RKKY) interaction between local moments mediated by conduction electrons. To incorporate this effect, we replace the temperature $\mathrm{T}$ in Eq.~\eqref{eq1} with an effective temperature,  $\rm T_{\rm eff} = \sqrt{T^2+T_W^2}$,  
where $\rm k_B T_W$ denotes the RKKY energy scale~\cite{V.K.C.Liang, J.Kästner, K.Fischer, Bikash_prb2020}. Using this $\rm T_{eff}$ in Eq.~\eqref{eq1}, the resistivity fits [as shown in the insets of Fig.~\ref{Fig1}(a) and (b)] yields $\rm T_K = 16\pm2$~K ($\rm I \parallel a$) and 9$\pm$2~K ($\rm I \parallel c$), with $\rm T_W =0.74\pm0.19 ~K~(I\parallel a)$ and $\rm T_W \approx 0~K~(I\parallel c)$, with spin quantum numbers $\rm S = 1.16\pm0.20$ and 0.89$\pm0.25$, respectively. These values are close to $\rm S=1$, consistent with a local-moment origin associated with W$^{4+}$ ions in a distorted Te$^{2-}$ environment (See SM~\cite{Note1}, Table-S1 for other fitting parameters).  

In Fig.~\ref{Fig1}(c–f), we present the temperature dependence of the magnetoresistance~(MR)~
for different current and magnetic field orientations (See SM~\cite{Note1}, Sec.~S4 for detailed analysis). Figure~\ref{Fig1}(c) shows the MR with $\rm I \parallel a$ for two magnetic field configurations, $\rm B \parallel a$ and $\rm B \parallel c$. In both cases, a broad low-temperature peak is observed, which originates from negative MR due to the suppression of Kondo spin-flip scattering under an external magnetic field. To quantify the anisotropy with respect to field direction, we define $ \delta({\rm MR}) = {\rm MR}(\rm B \parallel c) - {\rm MR}(\rm B \parallel a)$,
evaluated at $\rm B = 9$~T. The resulting temperature dependence of $\rm \delta({\rm MR})$ for $\rm I \parallel a$ is shown in Fig.~\ref{Fig1}(d). A pronounced downturn appears at low temperatures, demonstrating that the suppression of Kondo scattering is direction-dependent. For an isotropic Kondo scattering, $\delta({\rm MR})$ would remain featureless and resemble the clean reference sample S-1 (blue triangles), which shows no downturn in Fig.~\ref{Fig1}(d). The clear low-$\rm T$ downturn in S-3 therefore confirms anisotropic Kondo scattering.  

A similar analysis for $\rm I \parallel c$ is presented in Figs.~\ref{Fig1}(e) and \ref{Fig1}(f). Here too, negative MR is observed for both field orientations, consistent with suppression of Kondo spin-flip processes. However, the downturn of $\delta({\rm MR})$ is substantially smaller than that for $\rm I \parallel a$, reflecting weaker anisotropy. Together, these results establish that Kondo scattering in WTe$_2$ is not only disorder-dependent but also strongly anisotropic with respect to both current and magnetic field directions. This anisotropy is consistent with the underlying type-II Weyl dispersion, where the over-tilting of Weyl nodes renders the band structure highly directional~\cite{Soluyanov_nature15, Kamal_prb19}. Having established anisotropic Kondo screening in WTe$_2$, we now analyze the observation of spontaneous Hall effect in zero magnetic field.

{\it{ Spontaneous Hall effect (SHE):--}}~Figure~\ref{Fig2}(a) presents the measured transverse resistivity $\mathrm{\rho^{measured}_{xy}(B)}$ and longitudinal resistivity $\mathrm{\rho^{measured}_{xx}(B)}$ resistivity at T=2 K. The inevitable misalignment of the voltage contacts adds up a component of $\mathrm{\rho^{measured}_{xx}(B)}$ on $\mathrm{\rho^{measured}_{xy}(B)}$. To extract the actual Hall component, instead of direct antisymmetrizing, we adopted the following expression, $\mathrm{\rho_{xy}(B) = \rho_{xy}^{measured}(B) - [\rho_{xx}^{measured}(B) \times \rho_{xy}^{corr}(B=0)]}$. Here, $\mathrm{\rho_{xy}^{corr}(B=0)=\rho_{xy} / \rho_{xx}}$ is a correction factor measured at room temperature (300 K) and zero magnetic field~\cite{Dzsaber2,D.M.Kirschbaum}, defining the contact misalignment error. In this way, $\mathrm{\rho_{xy}(B)}$ preserves the exciting phenomena like SHE, the appearance of a B-symmetric Hall effect in nonmagnetic materials, which grows stronger as temperature decreases. 

The non-zero $\mathrm{\rho_{xy}(B=0)}$ in Fig.~\ref{Fig2}(b), indicates the presence of SHE. To extract the field dependence of the SHE conductivity ($\mathrm{\rho^{sp}_{xy}(B)}$), we used the two-band model fitting on $\mathrm{\rho_{xy}(B)}$ at high fields and extrapolated to zero field.  Figure~\ref{Fig2}(a) clearly shows an anomalous deviation at low magnetic field values (full range fitting is shown in the inset). The $\mathrm{\rho^{sp}_{xy}(B)}$ of S-3 at different temperatures shown in Fig.~\ref{Fig2}(c), are extracted by subtracting the two-band fit from $\mathrm{\rho_{xy}(B)}$. The SHE signal is symmetric with respect to the magnetic field reversal, confirming that it originates from the applied electric current and is not associated with the magnetic field. Generally, magnetic fields produce odd-$\rm B$ Hall responses. Notably, the SHE signal is significantly stronger in the most disordered sample, S-3, compared to S-1 and S-2 (see SM~\cite{Note1}, Sec.~S5). This highlights the important role of disorder-driven WKSM phase in enhancing zero-field spontaneous Hall transport in WTe$_2$, providing the first possible clue about its origin~\cite{Dzsaber2}.

Figure~\ref{Fig2}(d-f) shows the carrier density for all three samples, estimated via simultaneous two-band model fitting of $\rm \rho_{xx}(B)$ and $\rm \rho_{xy}(B)$~\cite{L.Wang, Y.Wang, ali2014large, D.Fu, R.Jha}. Our analysis reveals a systematic deviation from charge compensation with increasing disorder concentration. This trend is consistent with previous reports indicating that samples with higher RRR tend to be better charge-compensated~\cite{P.L.Cai, Z.Zhu, M.N.Ali2}. A more detailed analysis is presented in Sec.~S6 of the SM~\cite{Note1}. The fact that sample S-3 is not charge compensated indicates that the Fermi energy is slightly away from the Weyl node, as type-II Weyl semimetals are generally not charge compensated around the Weyl point energy. This is also seen in our theoretical calculation of electron and hole carrier density in an inversion broken type-II Weyl semimetal model, presented in Fig.~\ref{Fig4}(b).  The large imbalance between the electron and hole concentrations in S-3 [Fig.~\ref{Fig2}(f)] can also arise from a shift in the Fermi energy across other bands which are not associated with Weyl nodes. However, such trivial carrier pockets at the Fermi level in WTe$_2$ have not been reported.

\begin{figure}
\includegraphics[width=1\linewidth]{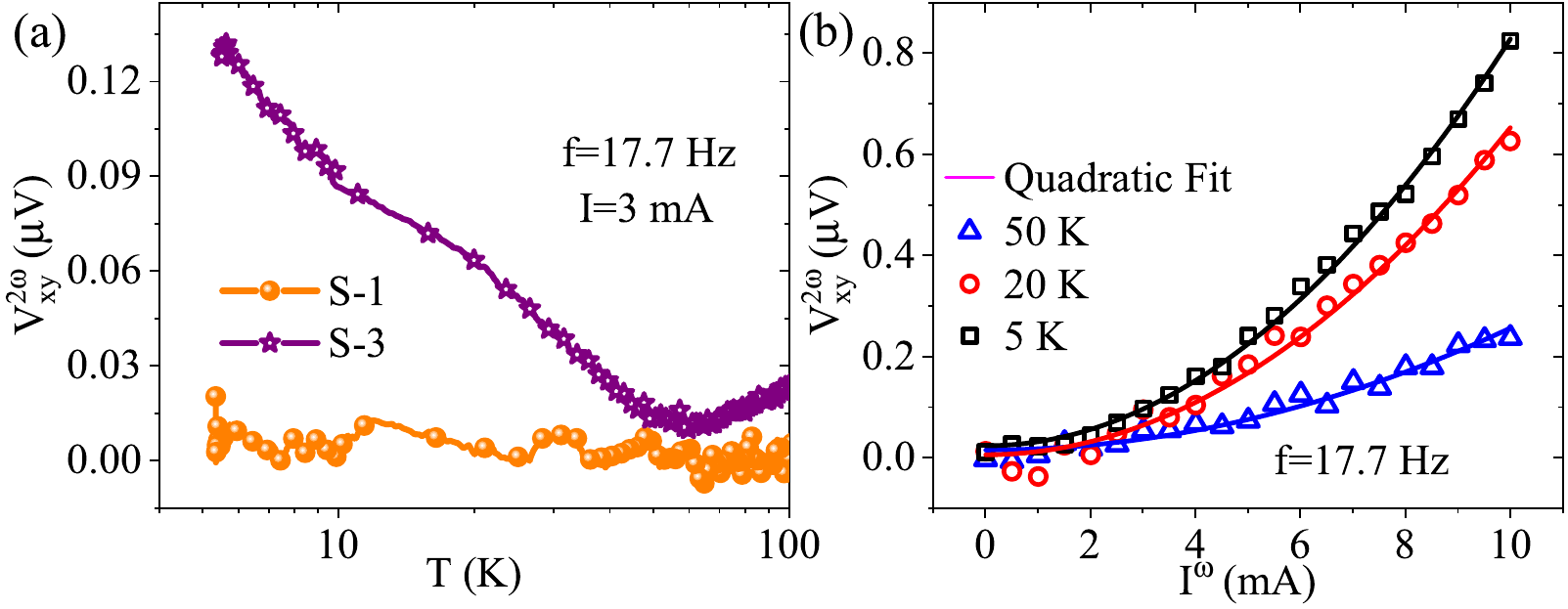}
\caption{{\bf Observation of BCD induced non-linear Hall effect.} (a) $\mathrm{2^{nd}}$ harmonic Hall signal ($\mathrm{V^{2\omega}_{xy}}$) of S-1 and S-3 with temperature at a fixed current 3 mA. (b) The quadratic behavior of the $\mathrm{V^{2\omega}_{xy}}$ with current (I$^\omega$) at different temperatures.}\label{Fig3}
\end{figure}

The appearance of spontaneous Hall response in WTe$_2$ is intriguing, since it is nonmagnetic and preserves time-reversal symmetry. As a consequence, a magnetization-driven anomalous Hall effect is forbidden in WTe$_2$. Surprisingly, we find that the spontaneous Hall response grows systematically with disorder, reaching its maximum in the Weyl–Kondo phase of WTe$_2$. In this regime, even weak electric fields can drive the system into a fully nonequilibrium regime~\cite{sur_25} where the distribution function becomes a non-perturbative function of the electric field. For an electric field applied along the $x$-direction, the nonequilibrium distribution function in the non-perturbative limit can be obtained by solving the Boltzmann transport equation. This is given by ${\rm g({ k})} = e^{\rm k_x/k_E} {\rm \int_{q_x} dq_x} e^{\rm -q_x/k_E} \rm (\partial_{q_x} f_0)$, with $\rm q_x\equiv k_x$, ${\rm k_E} =e \rm \tau E/\hbar$ where $\tau$ being the scattering time, and $\rm f_0$ is the equilibrium Fermi function (see Sec.~S7 of SM~\cite{Note1} for details). The ${\rm q_x/k_E}$ factor in the exponential makes the distribution function non-perturbative in nature. In this nonperturbative regime, the Berry curvature ($\rm \Omega_z$) contribution to the Hall current is given by, 
\bea
{\rm j_y^{sp}} = -\frac{e^2}{\hbar} \rm E_x \sum \int_{\bm k} \Omega_z g({\bm k})~.   
\eea 
To qualitatively capture the spontaneous Hall response in nonmagnetic WTe$_2$, we calculate $\rm j_y^{\rm sp}$ for a generic non-interacting tight-binding model of type-II Weyl semimetals with broken inversion symmetry~\cite{Soluyanov_nature15, Tewari_prb21}. See Sec.~S7 of SM~\cite{Note1} for the details of the model. Figure~\ref{Fig4}(a) shows the band structure of such a system with two pairs of Weyl nodes. In Fig.~\ref{Fig4}(b), we show the variation of the electron and hole carrier density with Fermi energy. Interestingly, we find that around the Weyl node (denoted by dashed horizontal lines), the system has a relatively larger charge imbalance~\footnote{Note that the band structure in Fig.~\ref{Fig4}(a) is plotted along a high-symmetry path with $k_z=0$. Along other paths with $k_z\neq 0$, the bands extend beyond $\varepsilon/t=3$. For clarity, we do not show these in Fig.~\ref{Fig4}(a), as our focus is on illustrating the pairs of Weyl nodes. As a result, the carrier density remains finite at $\varepsilon/t=3$ [see Fig.~\ref{Fig4}(b)].} This supports our experimental observation of the charge imbalance in S-3, where the Kondo interaction pins the Fermi level near the Weyl nodes. In Fig.~\ref{Fig4}(c), we present the calculated spontaneous Hall current along the $\rm \hat{y}$-direction for an electric field applied along $\rm \hat{x}$. 
This Berry curvature-driven fully nonequilibrium Hall response~\cite{sur_25,Dzsaber2} is the most likely origin of the observed spontaneous Hall response in our experiment.

\begin{figure}
\includegraphics[width=\linewidth]{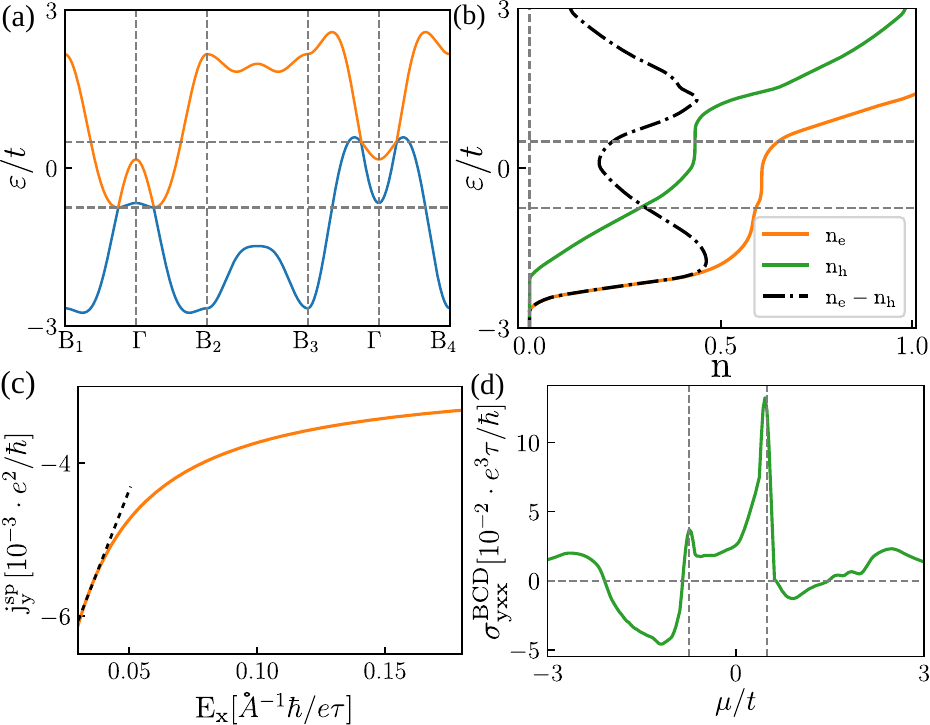}
\caption{{\bf Linear and nonlinear spontaneous Hall response in type-II Weyl semimetal.} (a) The band dispersion ($\varepsilon$ in units of hopping parameter $t$, see Sec.~S7 of SM~\cite{Note1}) of a generic time-reversal preserving type-II Weyl semimetal with broken inversion symmetry. (b) The number density in arbitrary units of the hole-like carrier and electron-like carriers. We observe that near the Weyl nodes (denoted by the dashed horizontal lines), the system is more charge decompensated. (c) Spontaneous Hall current versus applied electric field plot. The current $\rm j_y^{\rm sp}$ in the low field regime is linear in the applied electric field, giving rise to the spontaneous Hall response at the linear order (indicated by the dashed line). b) The BCD induced a nonlinear spontaneous Hall conductivity $\rm \sigma^{BCD}_{yxx}$ as a function of the chemical potential $\mu$. As evident from the plot, the nonlinear spontaneous Hall response peaks near the Weyl nodes ($\mu =0$), and it rapidly decreases with an increase in Fermi energy.}
\label{Fig4}
\end{figure}
We also performed second-harmonic Hall measurements ($\rm V^{2\omega}_{xy}$) across all samples. A finite $\rm V^{2\omega}_{xy}$ is observed in all three samples, with a markedly enhanced signal in S-3 compared to the cleaner sample S-1. Figure~\ref{Fig3}(b) shows $\rm V^{2\omega}_{xy}$ versus applied current $\rm I^\omega$ at various temperatures. 
As expected, $\rm V^{2\omega}_{xy}$ displays a clear quadratic dependence on $\rm I^\omega$. Moreover, the second-order nonlinear longitudinal response is found to be negligible, ruling out longitudinal signal mixing. 
The strong enhancement of $\rm V^ {2\omega}_{xy}$ below 50 K in S-3 indicates a robust Berry curvature dipole-driven Hall response, which gets suppressed at higher temperatures.

The second-order nonlinear Hall response induced by the Berry curvature dipole (BCD) has also been previously observed in WTe$_2$ thin films~\cite{Kang_nm19, 
Pablo_NP18}. For an electric field along $\rm \hat{x}$, the conventional BCD-driven Hall current 
is given by
\bea
\rm j^{BCD}_y &=& \rm \sigma_{yxx}^{\rm BCD} E_x = - \frac{e^3 \tau}{\hbar} \Lambda_{zx} E_x, \nn \\
\rm \Lambda_{zx} &=& \rm -  \int_{\bm{k}} \Omega_z v_x \frac{\partial f_0}{\partial \varepsilon},
\eea
where $\rm \Lambda_{zx}$ denotes the BCD.
Model calculations for a type-II Weyl semimetal [Fig.~\ref{Fig4}(d)] show that the BCD conductivity peaks near the Weyl node and vanishes when the Fermi level is far away from the Weyl node.  This behavior is consistent with our observations: in S-3, Kondo interactions pin the Fermi level close to the Weyl nodes, leading to a pronounced nonlinear Hall response. In contrast, if the Fermi level is far away from the Weyl node in S-1, the resulting $\rm V^{2\omega}_{xy}$ singal is expected to be much smaller. 


{\it { Conclusion:--}} We identify WTe$_2$ as a promising Weyl–Kondo semimetal candidate situated near quantum criticality, where moderate disorder tuning provides access to the WKSM ground state. The emergence of anisotropic Kondo screening, together with disorder-enhanced spontaneous Hall effects, indicates that Kondo interactions pin the Fermi level near the Weyl nodes. In this regime, the small Fermi surface renders the system highly susceptible to fully nonequilibrium transport: even weak electric fields can drive it beyond the perturbative limit, giving rise to a Berry-curvature–driven spontaneous Hall response at linear order. At the same time, the Berry curvature dipole generates a pronounced nonlinear Hall effect, maximized when the Fermi level is close to the Weyl nodes. These results demonstrate a strong interplay between Kondo physics and the band topology of a type-II Weyl semimetal. While the full theoretical description of disordered type-II WKSMs is still incomplete~\cite{Xie_prb17,Matthew_prb17}, our work establishes WTe$_2$ as a promising platform to investigate disorder-induced correlated topology in nonmagnetic Weyl semimetals.

\section{acknowledgments} The authors acknowledge IIT Kanpur, the Department of Science and Technology (DST), India, [Order No. DST/NM/TUE/QM-06/2019 (G)] and DST-FIST (DST/PHY/2022 523) for financial support. SD is supported by the Prime Minister's Research Fellowship, Ministry of Education, Government of India.

\bibliography{refs}
\end{document}